# On the effectiveness of imposing restrictive measures in a graded Self-Organized Criticality epidemic spread model: The case of COVID-19

Y. Contoyiannis[1], S.G. Stavrinides[2], M.P. Hanias[3], M. Kampitakis[4], P. Papadopoulos[1], S. Potirakis[1]

[1]Department of Electric-Electronics Engineering, University of West Attica, Athens, Greece. (yiaconto@uniwa.gr, ppapadop@uniwa.gr, spoti@uniwa.gr)

[2]School of Science and Technology, International Hellenic University, Thessaloniki, Greece. (s.stavrinides@ihu.edu.gr)

[3]Physics Department, International Hellenic University, Kavala, Greece (mhanias@physics.ihu.gr)

[4]Major Network Installations Dept, Hellenic Electricity Distribution Network Operator SA, Athens, Greece (m.kampitakis@deddie.gr)

**Abstract:** The scope of this work is to serve as a guiding tool against subjective estimations on real pandemic situations (mainly due to the inability to acquire objective real data over whole populations). The previously introduced model of closed self-organized criticality (SOC), is adapted in the case of a virus-induced epidemic. In this version this physical model can distinguish the virus spread according to the virus aggressiveness. The study presented, highlights the critical value of virus density over a population. For low values of the initial virus density – lower than the critical value – it is proved that the virus-diffusion behavior is safe and quantitatively similar to usual real epidemical data. However, it is revealed that very close to the critical point, the critical slowing-down (CSD) phenomenon, introduced by the theory of critical phenomena, emerges, leading to a tremendous increase of both the percentage of active carriers and the duration of the epidemic. A behavior of the epidemic obeying to a second order phase transition, also occurs. For virus density values higher than the critical value, the epidemic duration becomes extremely prolonged. Additionally, the effect of the closed system population size revealed interesting properties. All these results, together with an investigation of the effectiveness of applying physical contact restriction measures, document scientifically their worthiness, while they also demonstrate the limits for which *herd immunity* holds safely. Finally, the model has been compared against real epidemic data in the case of Greece, which imposed restrictive measures consistently and in time.

## 1. Introduction

The study of viruses' behavior, utilizing multidisciplinary frameworks, has revealed the existence of universal patterns. Viruses demonstrate both high instability and adaptability due to their elevated mutation rates, further leading to diversified characteristic behaviors including their level of aggressiveness. Thus, the resulting dynamics involve nonlinear phenomena, criticality (tipping points), and self-organization.

Models on epidemic spread, in terms of the virus self-organized criticality (SOC) [1], have been developed in the recent past [2, 3, 4]. Formulating a model is based in principle, on the number of factors, one must consider, as well as the way these factors would be quantified and represented within the model. These factors may additionally refer to the rules instituted by the authorities, in order to prevent the spread of an epidemic.

A lot of discussion and controversy over the issue of restriction measures imposed by authorities takes place worldwide, on occasion of COVID-19 pandemics. The scope of this work is to investigate the effect of protection measures on the spread of a virus over a population, considering this as a self-organized system; aspiring to serve as a guiding tool against subjective estimations on the real pandemic situation (mainly due to the inability to acquire objective real data over whole populations).

The methodology followed herein, as well as the introduced model-quantities, have already been proposed and utilized in the case of self-organized systems [1], while diffusion rules similar to the sandpile model of BTW [5], have been imposed. This was further adapted to a virus-exposed population-system model, including the capability to consider various levels of virus aggressiveness and allowed physical contacts. The importance of critical density $\rho_c$ (defined in the following section) is introduced, together with the critical slowing-down (CSD) phenomenon, emerging on the critical point; both coming from the theory of critical phenomena [6]. It is noted that the CSD phenomenon has been thoroughly studied in epidemiology [7, 8].

An observable measure is the percentage of the infected people during an epidemic (always estimated, never measured). Such realistic percentages in the cases of normal virus diseases, are between 0.5%-1.5%; not excluding of course the possibility for an epidemic to exceed these limits. In this work, an investigation on whether a self-organization model can provide such realistic results, appears together with an estimation of the epidemics duration. This investigation is further enriched by introducing in the model physical contact restriction (epidemic control) rules, something that happens in many countries in the era of COVID-19. Finally, interesting results document the limits of the *herd immunity* approach and the usefulness and effectiveness of the imposition of contact restriction rules. These results are further compared to real world data in the case of Greece, while restriction measures were imposed.

## 2. Method - The proposed model

The two fundamental quantities in the models of self-organized criticality (SOC) are the control parameter and the order parameter. Attempting to introduce such a model in the case of a virus-exposed population-system, these two parameters are defined as follows:

- The control parameter is defined as the virus density $\rho = \frac{Q}{P}$, where $Q$ is the number of the total virus units, which in the rest of this paper would be called as the virus charge ($Q$), and $P$ is the overall population, over which this virus charge has been distributed. In the proposed model, population ($P$) would be considered as an ($L \times L$) lattice, where each site corresponds to a person; therefore, $P=L \times L$. It is noted that the initial virus charge $Q_o$, is always distributed randomly over the lattice sites (namely the population).
- The percentage of virus carriers, capable of transmitting the virus ($C$), over the total population ($P$) defines the order parameter $M = \frac{C}{P}$. The sites corresponding to those carriers are named active sites. In the model introduced and investigated hereby, it is $M(\tau)$ the quantity that finally represents the extent of the spread of a virus disease, since this magnitude enunciates the percentage of the active carriers (including the asymptomatic ones). As a result, by following the evolution of $M(\tau)$ one may have a good quantitative description about the evolution of the epidemic. It is noted that the proposed model has the potential to demarcate active from inactive carries of the virus.

Regarding the virus unit, this is defined as the mean maximum number of viruses propagated in each contact, between two people. It is apparent that the so called in this work virus unit represents a large number of (biological) viruses and it is an object of medical designation. This way a correspondence between the initial BTW model and the proposed hereby, could be established, i.e. a virus unit corresponding to a grain of sandpile. In the BTW model a site is active when the number of grains is $n \geq 4$. In the proposed model, regarding the virus case, things are more complex in the sense that a discrimination between an active and an inactive carrier is introduced. Due to the large number of (biological) viruses comprising a single virus unit (with reference to the model), a quantization of the virus charge that is transferred from one lattice site (carrier) to its four neighboring sites, could be performed. This quantization is performed according to the number of virus units that the carrier possesses when is active. Therefore, in this model a carrier could be active even for $n \leq 4$ and the virus charge that could be transferred from one lattice site to another, is defined by the relation $n/4$, where $n=1,2,3,4$. Thus, the transferred virus charge could be either 1 (for $n=4$) or a fractional number, belonging to (1/4, 1/2, 1/3) – for $n=1,2,3$. In the case that $n=1$, all the carriers are active, spreading aggressively the virus over a population; this appears to be one of the qualitative differences of COVID-19, compared to other flu-viruses. It should be noted that this is a major differentiation compared to the original BTW model, where the only active case that exists is for $n=4$, meaning that the active carrier is allowed to transfer only one charge (due to the transfer mechanism of the phenomenon that it describes). As a result, the rules of the virus charge diffusion model over the whole lattice, prescribes that every active site can transfer $n/4$ virus units to each of its four neighbors. It is obvious that this discrete (quantized) model has a coarse graining character, since it is able to "see" details according to the quantization; inherently including the aggressiveness of the virus studied, though. Finally, it should be mentioned that this nearest-neighbor approach, adopted in this model, is indeed appropriate for describing the spread of a viral epidemic, since it can simulate virus propagation through any kind of close contact between people, characterized by a short radius range.

Implementing the model, for each sweep of the lattice a time unit is mapped, correspondingly. This way a timeseries of the order parameter $M(\tau)$ can be created. There is a critical value $\rho_c$ for the virus density, for which if the initial density (in the beginning of diffusion procedure) is less than this value ($\rho<\rho_c$), then the final order parameter $M(\tau)$ value becomes zero, after a relaxation time period $\Delta\tau$. This means that after this period no active sites (people who can propagate the virus) there exist, within the whole population, although there may be virus carriers (inactive). Therefore, the studying the model's behavior at the critical point is noteworthy interesting.

## 3. Results

An example of a closed system with a population of $P$=14400 people has been considered. For the proposed model, this was translated to a square lattice of 120x120 ($L$=120). In order to have a realistic and timely case, $n$ was set to $n$=1, meaning that all virus carriers were considered to be active; thus, the option of studying an aggressive virus, like COVID-19, was chosen.

Initially, and in order to find the critical value $\rho_c$ of the virus density, the order parameter $M(\tau)$ timeseries for various initial virus density values $\rho$, was generated. In Fig. 1 the dependence of relaxation time $\Delta\tau$ on the virus density $\rho$ is presented. The highest value of the virus density, for which the final order parameter value $M(\tau)$ becomes zero, determines the critical value $\rho_c$ of the virus density. This value was found to be $\rho_c$=0.536 (for the 120x120 lattice). In addition, the change of the virus charge (i.e. the total virus units) was calculated to be $Q$=0.536·14400=7718 virus units.

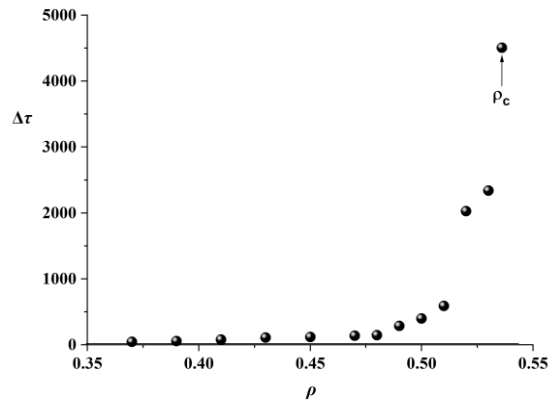

**Figure 1:** Relaxation time $\Delta\tau$ vs virus density $\rho$, for $n$=1. During the CSD phenomenon, relaxation time $\Delta\tau$ steeply increases when approaching close to the critical point [6].

It is apparent from Fig. 1 that in this case, relaxation time goes to an unusual divergence at the critical point. This is a well-known property described in the theory of critical phenomena, namely the critical slowing-down (CSD) phenomenon; during this phenomenon, getting very close to the critical value of a variable, the studied magnitudes are evolving with a rate that slows-down, leading to high relaxation times. Thus, as we reach the critical virus density time duration of the epidemic spread becomes many times greater.

### A. Initial virus density $\rho \leq \rho_c$

As expected from the theory of critical phenomena, in the case of initial virus densities $\rho$ below the critical value, time durations of the epidemic spread become significantly small. Therefore, investigation of the characteristic scenario (when $\rho = \rho_c$) for the temporal evolution of order parameter $M(\tau)$, where the greatest relaxation time emerges, is presented in Fig. 2. Looking at this graph it comes up that for a relaxation time $\Delta\tau$=4505, the order parameter $M(\tau)$ gets a zero value, meaning that the epidemic has ended. It is noted that in the central region of relaxation time (1000 <$\tau$ <2000 in Fig. 2), the percentage of the active carriers drops to a mere mean value of 2.2%, which is a very low percentage for active carriers, in any case.

Actually, one of the goals of this model is the fact that it studies the epidemic spread after this has reached its climax. In the model the climax is considered as the initial point of the simulations. Consequent to the above, in the example appearing in Fig. 2, the initial percentage of the active carriers M($\tau$) has been considered to be almost 10% and it appears to reduce very quickly. Therefore, the focus of this study is on the most interesting part of the graph, that of the fluctuating process towards the disappearance of the epidemic for $\rho \leq \rho_c$.

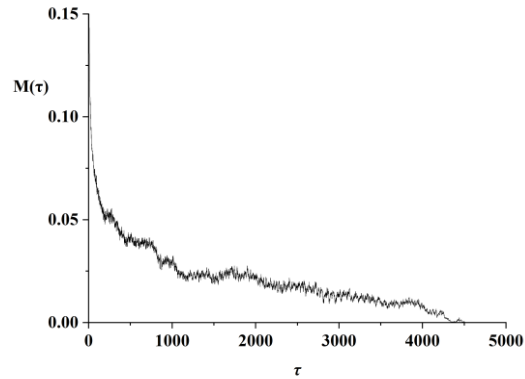

**Figure 2:** The order parameter $M(\tau)$ vs. time $\tau$ exactly on the critical point, where $\rho = \rho_c$. The order parameter $M(\tau)$ represents the extent of the spread of a virus disease, over the population.

## B. Initial virus density $\rho > \rho_c$

Beginning the simulation from virus densities higher than the critical value a different behavior emerges. The order parameter $M(\tau)$ initially increases, saturating to an almost constant value, for higher relaxation times, finally leading to infinite relaxation times; thus, virus spread never reaches an end (practically the spread lasts extremely long times). This behavior is illustrated in Fig. 3; the evolution of the order parameter $M(\tau)$, for a virus density slightly higher than its critical value, namely for $\rho=0.600$ (>0.536), is presented. It is apparent from this figure that next to the practically infinite duration of the virus spreading, it is confirmed that even small variations above the critical virus density lead to increased percentage of active carriers – in this case it is almost 22.5%, ten times more than in the case of the critical value. Thus, the critical value $\rho_c$ emerges as the borderline between usual, normal spread of the virus and an unusual, aggressive spread demonstrating with high duration times.

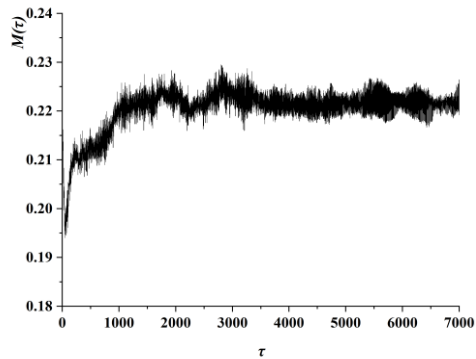

**Figure 3**: A segment of 7000 points of the graph the order parameter $M(\tau)$ vs. time $\tau$, for $\rho=0.6 > \rho_c=0.536$ (after the critical point). As seen in the graph the disease appears not to reach an ending point.

In order to get an insight of the demonstrated phenomenon, the plot of the order parameter $M$ vs. the virus density $\rho$, is presented. The resulting graph in Fig. 4(a), obviously shows a pattern that is characteristic in nonlinear dynamics (similar graphs contributing to the investigation of the nonlinear behavior of the model appears in [9], as well); this pattern is the well-known "devil's staircase", clearly hinting for a nonlinear phenomenon [10, 11].

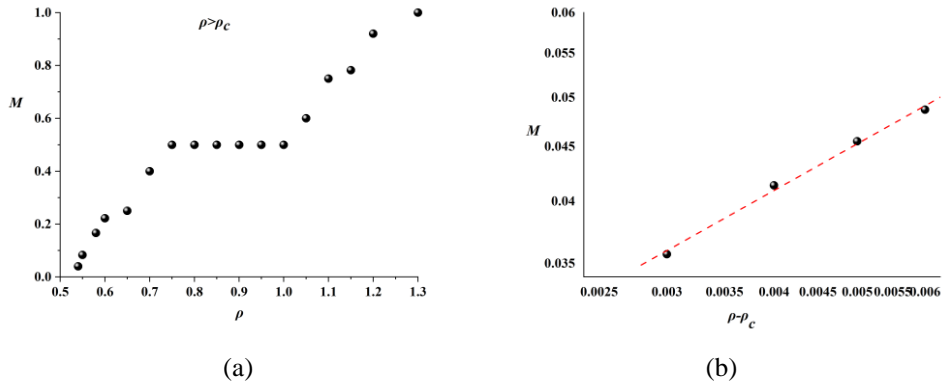

(a)  (b)

**Figure 4:** (a) The order parameter dependence on the virus density in the case $\rho=0.6 > \rho_c=0.536$. A characteristic quite large plateau appears for $M(\tau)=0.5$. (b) Illustration of the power law that clearly holds very close to critical point. The calculated value for the characteristic exponent is $\beta=0.45$.

According to the theory of second order phase transitions, it is expected that close to the critical point a power law of the form $M\sim(|\rho-\rho_c|)^{\beta}$, should hold. In order to investigate this feature, the corresponding graph was plotted, and appears in Fig. 4 (b). Indeed, getting close to the critical virus density, for $\rho>\rho_c$, such a power law clearly holds, while the characteristic exponent $\beta$ of this power law was calculated to be $\beta=0.45$.

### C. Factors influencing the duration of an epidemic

As already described above, in the model introduced and investigated hereby, the quantity that finally represents the extent of the spread of a virus disease is the order parameter $M(\tau)$, therefore, by following its temporal evolution, one may have a good quantitative description about the evolution of an epidemic. As already discussed, the proposed self-organized model seems to demonstrate a noteworthy behavior for a virus density equal to its critical value $\rho_c$, due to the CSD phenomenon, further leading to spectacularly increased relaxation times (high durations of the epidemic).

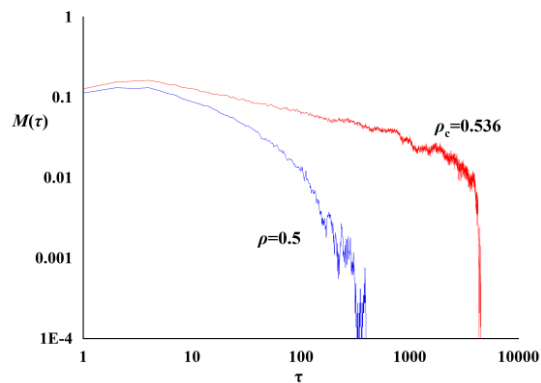

**Figure 5:** A comparative graph, illustrating the temporal evolution of the order parameter $M(\tau)$ exactly at the critical point $\rho_c=0.536$ (red line) and $\rho=0.500<\rho_c$ (blue line). It is apparent that under the critical value, where the effect of the CSD phenomenon does not exist, the relaxation time (epidemic duration) appears to be extremely reduced.

In order to stress and illustrate this feature of the model, the order parameter temporal evolution $M(\tau)$ appears in the double graph in Fig. 5, for $\rho=\rho_c=0.536$ (red line) and for $\rho=0.500<\rho_c$ (blue line), close to the critical point, but out of the region where the CSD phenomenon takes place. In both cases the model was implemented for a 120x120 lattice. It is apparent from the graph that for $\rho\leq\rho_c$ relaxation time drastically drops; in the studied cases, from $\Delta\tau_c=4505$ to $\Delta\tau_{\rho=0.5}=398$. This enormous difference permits us to describe quantitatively, through relaxation time, how close to the "dangerous" critical point the system lays, without the requirement to define the time measuring units.

*TABLE I*
Comparison of the model results for three different populations

| Population | Epidemic duration (Δτ) | <M>, after τ=1000 | Δτ·<M> |
|---|---|---|---|
| P = 120x120 = 14400 | 4505 | 1.31% | 59 |
| P = 200x200 = 40000 | 7507 | 0.35% | 26 |
| P = 270x270 = 72900 | 9061 | 0.22% | 20 |

Likewise, and in order to check the influence of the size of the population (in the model, the size of the lattice) on the temporal escalation of the virus disease, the temporal evolution of the order parameter $M(\tau)$ was calculated, in two more lattice cases: a 200x200 lattice, i.e. a population of 40000; and a 270x270 lattice, i.e. a population of 72900. The results of this study are presented in Table I. For the shake of clarity, in Fig. 6 the graphs of $M(\tau)$ vs. $\tau$ for all three cases of population, $P=14400$ (red line), $P=40000$ (green line) and $P=72900$ (blue line), appear. The information that can be mined from this comparative illustration are quite interesting and intriguing and are apposed in the following list:

- The epidemic duration appears to be higher with increasing populations, as naturally expected.
- The range of the fluctuations appears to be larger for smaller populations.
- The average value of the percentage of the diseased population $M(\tau)$ is higher in the case of the smallest population.
- No conservative magnitudes have been observed during the evolution of the phenomenon. On the contrary, even the area underneath each graph in Fig. 6, which is analog to the corresponding product $\Delta\tau\cdot<M>$ (last column in Table I), appears to decrease with increasing population.

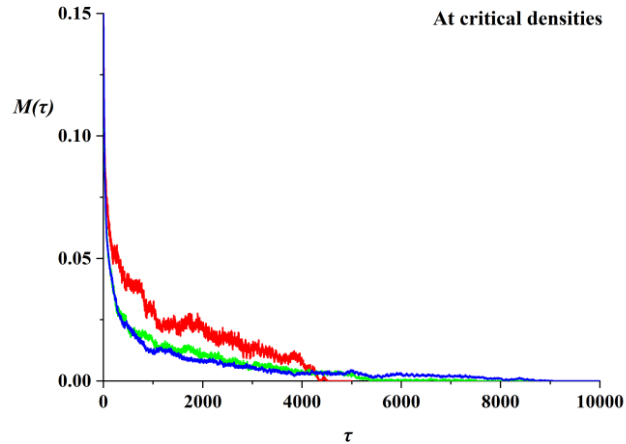

**Figure 6:** A comparative diagram of the temporal evolution of the order parameter M(τ), which represents the percentage of the ill population, vs. time τ, is presented, for three populations of different size. The plots of all three cases were calculated at the critical virus density ρc. The red plot corresponds to a population of 14400, while the green plot to 40000 people and the blue color to 72900 people.

The last result comparative-presentation regards the effects of the virus aggressiveness on the model behavior. As deducted from the above presented results, the metric that defines the boundaries between a usual (normal) virus spread within a population, and an unusual (with extremely high rate) spread, is that of the critical virus density $\rho_c$. On the other hand, in the proposed hereby model the aggressiveness of a virus has been quantified in 4 discrete levels, according to the number of virus units that the carrier possesses, when is active ($n=1,2,3,4$). For $n=1$ all the carriers within a population are active (carriers propagating the virus), while for $n=4$ only some of the carries are active. It is apparent that the aggressiveness of the virus, influences the value of the critical point for the epidemic. In Table II the values of the critical virus density $\rho_c$. according to the virus charge are presented, in the case of a population of 14400 (120x120).

*TABLE II*
The calculated $\rho_c$ according to the virus aggressiveness for a population of 14400

| *n* | *n/4* | *ρc* |
|---|---|---|
| 1 | 1/4 | 0.536 |
| 2 | 1/2 | 1.065 |
| 3 | 1/3 | 1.848 |
| 4 | 1 | 2.1420 |

It is apparent from this table that the width of safe initial conditions ($\rho > \rho_c$) is enhanced for less aggressive virus diseases. In specific, in the case of an extremely aggressive virus (as in the case of COVID-19) the value of the critical virus density ($\rho_c$) becomes 4 times less (the width of the safe range of initial conditions for $\rho$, shrunk), compared to that of a normal virus disease (for instance a typical flu).

As a result, in the case of less dangerous viruses reaching the critical virus density is more difficult, because increased virus charge is demanded, resulting into shorter epidemic duration. On the contrary, in the case of more dangerous viruses reaching the critical point becomes easier, due to the reduced virus charge demanded. These results further confirm the consistency of the proposed model and allows for grading the expected risk.

**D. Results when restrictive measures are imposed**

The role of the authorities that have the will to protect their subjects, is to intervene to an epidemic's spread within their territory, by utilizing various tools. Some of those tools aim to the reduction of contact and interaction between their subjects, finally intending to intense and severe reduction of the virus spread rate.

In order to represent and introduce into our model, the imposed restrictions in human contact and interactions, the option of the model to define the bonds between the lattice sites, was utilized. Thus, three out of the four nearest neighboring site-bonds in the lattice were disabled, in a random way. This way, the active sites, carriers capable of transmitting the virus (infecting one neighbor) demonstrate a reduced infection capability.

In this case the reduction of the contact was chosen to be 75% (1 out of the 4 bonds remains active) – something that happens indeed in nations with disciplined citizens that take such measures. This 75% reduction in physical contacts was embodied within the model by allowing active nodes (carriers) to contaminate only one out of four immediate neighbors; which was decided by utilizing a random-bit generator (RBG). In specific, each neighbor belonged to one of the four equal subsets that the [0,1] set was divided to; the RBG decided which subset would be the active one, thus, which one of the neighbors would be in contact to the active carrier. Due to the lattice symmetry (in this approach), all four neighbors were subjects to round shifts, so that each neighbor could get values from all four subsets. This way 16 combinations there existed, and the time length for the epidemic duration was calculated for each combination. The combination providing the value closest to the average, is the one considered as the epidemic duration, under the specific circumstances.

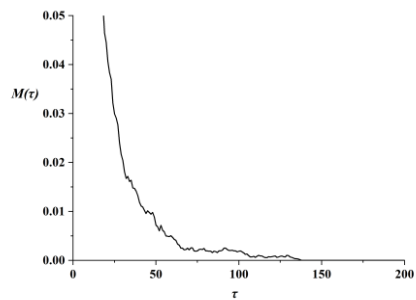

**Figure 7:** The temporal evolution of the percentage of ill population, after imposing restricting measures. In the proposed model the measures were translated into the isolation of 3 out of 4 nearest neighbors. In this case the duration of epidemic falls by 33 times, for the same virus density.

Application of this indicative case of restrictions over the previously studied 120x120 lattice (14400 people), in the case of an initial virus density $\rho=\rho_c$, exactly on the value of the critical virus density, is expected to lead to some kind of deceleration in the spread of the disease. In Fig. 7 the corresponding temporal evolution of the order parameter $M(\tau)$ is illustrated, not only fully confirming these expectations, but also revealing a spectacular reduction of the relaxation time ($\tau$) by at least one order of magnitude. The duration of the epidemic now appears to become $\Delta\tau=143$, almost 33 times less than that demonstrated in the case of lack of any restrictions of contact measures.

For the shake of clarity, the obtained results in both cases, without and with restrictions imposed, are presented in Table III.

*TABLE III*
Comparison of the model results with and without restrictions on human contact introduced

|  | *Epidemic maximal duration $\Delta\tau_c$* | *Percentage of diseased people* |
|---|---|---|
| *Without restrictions* | 4505 | 2.2 |
| *With restrictions* | 143 | 0.22 |

All the above hint for the fact the evolution of epidemics, according to the proposed model (also verified by reality) is a strongly nonlinear phenomenon. In addition, the huge reduction effect on the epidemic spread, of any restriction measures in the contact and interaction between the people, emerges; thus, imposing isolation measures seems to become an imperative necessity.

### E. The paradigm of Greece

In an attempt to evaluate the validity of the proposed model in the case of timely and drastic restrictive measures, the real COVID-19 epidemic data in Greece, as these were registered for the period between April 8th to 23rd, were compared to the numerically calculated ones (described in the previous subsection), regarding the case of imposing restriction measures leading to a reduction of physical contacts by 75% (Fig. 7).

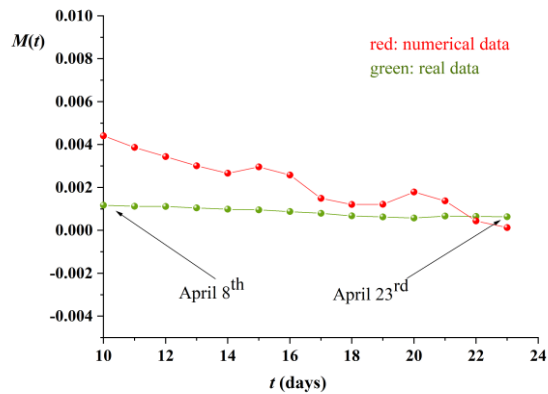

**Figure 8:** The percentage of active carriers in Greece for the period between April 8th to April 23rd. Real data (in green) according to the (Hellenic) National Public Health Organization are compared to the numerical results (in red) in the case described in the previous subsection (Fig. 7). The convergence between real and calculated data to a less than 0.5% is obvious, for the proposed scaling.

It is widely accepted that due to the lack of COVID-19-infection-test-results over the whole population that would result into the inclusion of all the cases, even the asymptomatic ones, the registered COVID-19 epidemic data is estimated to be about 10 times less than the real situation. Therefore, for comparison reasons the percentage of the infected (active) population was calculated to a population of 1 000 000 rather than 10 000 000, which is almost the population in Greece. Additionally, numerical data was processed in the frame of algorithmic time, with a time-step being one lattice scan, while for the real data a time-step was considered to be one day. In order to achieve the same time-length in both cases, the average of six numerical lattice-scans was mapped to one day. This way, a timeseries of the percentage of the active carriers $M(t)$, according to the model (in red), was derived and it is presented in Fig. 8, together with the real daily data (in green), according to the (Hellenic) National Public Health Organization (publicly available in https://eody.gov.gr/epidimiologika-statistika-dedomena/imerisies-ektheseis-covid-19/ in Greek and https://eody.gov.gr/en/current-state-of-covid-19-outbreak-in-greece-and-timeline-of-key-containment-events/ in English-abstractive). Both plots appear in a common timescale.

Thus, in the case of Greece and while restriction measures were in force, the model predicted a percentage of active carriers $M(t)$ being less than 0.5% by April 23$^{rd}$, something confirmed by the real epidemic data. The convergence between real and calculated data is obvious for the proposed scaling and provides with a consistent description of the COVID-19 epidemic by the proposed model, in relaxation phases.

## 4. Discussion and Concluding Remarks

In this work the study of the evolution of epidemics, based on the closed model of self-organized criticality (SOC), is presented. In this model the virus-system is considered in both cases: without any intervention and with interventions, which are considered to affect structural elements of the model. The study considered the climax as the initial point of the simulations, since relaxation of an epidemic is currently of interest. Future work would focus on the model's necessary modifications, so that it would be capable of describing all three stages, accession the climax and relaxation.

The main conclusion derived by this study is that if the virus-system is allowed to be freely organized by itself, then the risk of having extremely enhanced and prolonged disease spread, becomes high. In addition, a metric originating from the theory of critical phenomena, that of the critical virus density $\rho_c$ emerges as an important factor. For a specific population, if the initial virus density $\rho$ is greater than its critical value, then the epidemic never ends, and the infected population percentage remains at almost constant levels, depending on the density value. On the contrary, for initial virus density $\rho$ lower than its critical value, the epidemic ends relatively soon. It should be noted that in the latter case, and for as long as the epidemic is in the phase of spreading, the emerging results agree qualitatively with real-world data. This is a result not expected at all, since what sounds reasonable is that self-organization without any imposed control would lead to fully uncontrolled and dangerous occasions (regarding the disease spread), in any case. It is noted that getting close (from below) to the critical value of the virus density, the duration of the epidemic becomes peculiarly increased, due to the classic in the theory of criticality, critical slowing-down (CSD) phenomenon; thus the percentage of infected, active population could eject to almost an order of magnitude higher values. A comparison of the effects on the safety limits for the virus spread, on the aggressiveness of the virus, shows a drastic reduction of the critical virus density value $\rho_c$; thus, an abridgement of the safety limits regarding the virus spread is evident.

In addition, the effects of imposed restriction measures by the authorities, were studied within the frame of the proposed SOC model. These measures were considered to bare a change on the properties of the model's lattice sites, thus on the active carriers' capability to transfer their virus charge to their neighbors, namely a 75% reduction on the contact and virus-transfer capability for each site (person). The results clearly show a spectacular reduction of both the temporal duration of the epidemic and the percentage of the ill population. These results have been compared to real data in the case of Greece and they provided with a proof of concept for the proposed model which demonstrates a consistent description of the COVID-19 epidemic in relaxation phases.

Finally, an investigation of the role of the size of the population of the closed self-organized model revealed interesting properties, which should be considered by policy and decision makers. It appears that although closed systems of smaller populations are demonstrating shorter disease-durations, they exhibit large fluctuations and a higher average in the diseased population.

Concluding, it seems that self-organization, a mechanism applied in nature, as well as in artificial (social, economic etc.) networks, appears to have application in the description of virus-caused epidemics, as well. The proposed SOC model proves that the epidemics get out of control in the case of allowing the percentage of the infected population to get beyond a threshold. Therefore, the proposed by many sides (scientific, political, social etc.) solution of the *herd immunity* approach, is safe as long as these limitations hold. Since in real-world, human population systems, this threshold (i.e. the critical value of the virus density) dividing the safe side from the unsafe, is unknown and difficult to be calculated or estimated, it is apparent that significant risk lurks. In addition, variations of this threshold appear to be prone to virus characteristics related to its aggressiveness; apparently mutations could further complicate any possible estimation. Consequently, restricting measures imposed by the authorities are emerging as an imperative solution to the right direction; and in this study, the solution of contact restrictions introduced in the model, showed that it leads to extremely reduced temporal durations of the virus disease. Additionally, fragmenting and isolating population sets are approaches clearly implied by the closed form of the studied, self-organizing model. Although this fragmentation to small population sets appears to lead to smaller durations of the epidemic, it seems that the average percentage of diseased population is higher, being accompanied by intense fluctuations; meaning that decision makers should take all these into account, when deciding the quarantine fragmentation sets and the restriction rules. Thus, a scientifically documented coarse roadmap of the measures that policy makers should apply and their questionings and limitations, emerges.

During these difficult times of the COVID-19 pandemic, controversy over the proper approach in managing this calamity holds, worldwide. European countries and USA adopted different approaches, ranging from *herd immunity* and limited (Sweden, Britain) or late (Italy, Spain, USA) restriction measures to timely and total impose of a set drastic restrictions (Greece). The case of Greece, which successfully imposed radical restrictive measures in time, on the almost closed system at a national level, restricting human contact and direct interaction by about 80%, appears to be a verifying real-world example of the above described model and its results.

**Author's Contribution**

Y.C., S.G.S. and M.P.H. wrote the main manuscript and prepared all the figures. Y.C., M.K., P.P. and S.P wrote and run the code of the model. All authors actively cooperated in developing the proposed model, revised and checked the code at all stages, and reviewed the manuscript.